\documentclass[11pt,a4paper]{article}

\usepackage[utf8]{inputenc}
\usepackage[T1]{fontenc}
\usepackage{amsmath, amssymb, amsthm}
\usepackage{graphicx}
\usepackage{authblk}
\usepackage{hyperref}
\usepackage{geometry}
\usepackage[normalem]{ulem}

\usepackage[dvipsnames]{xcolor}

\title{Competition between evidence-based and alternative medical paradigms in non-infectious diseases}

\author[1]{Hermes Benito-Andr\'es}
\author[2]{Alfonso de Miguel-Arribas}
\author[1]{Carlos Gracia-L\'azaro}

\affil[1]{School of Architecture and Technology, Universidad San Jorge (USJ), Autovía Mudéjar, km. 299, 50830, Villanueva de Gállego, Zaragoza, Spain.}
\affil[2]{Zaragoza Logistics Center (ZLC), Av. de Ranillas 5, 50018 Zaragoza, Spain}

\date{\today}

\begin{document}

\include{vector}

\maketitle
\begin{abstract}
Standard epidemiological models coupled with evolutionary game theory (e.g., vaccination games) typically rely on the free-rider dilemma inherent to infectious diseases. In this paper, we present a coupled behavior-disease model where the underlying disease is non-infectious, and transmission occurs strictly at the cultural level. We model the competition between two medical paradigms, namely evidence-based medicine ($M$) and alternative medicine ($A$), using replicator dynamics driven by treatment-specific social-amplification coefficients. We show analytically that the nonlinearity of the coupled system arises from social feedback rather than from the underlying biological dynamics. Crucially, the stability analysis reveals that the system lacks isolated fixed points; instead, it converges to a one-dimensional continuous manifold of non-hyperbolic neutral equilibria, generating persistent path dependence in the cultural composition. Furthermore, we derive an explicit constant of motion that constrains each trajectory to a distinct invariant level set. Restricting this invariant to the equilibrium manifold yields a strictly monotonic selection function, uniquely determining the equilibrium selected from any given initial state. Our findings show that derivative-based cultural feedback is sufficient to generate a continuum of coexisting health paradigms, while asymmetric social amplification modifies this selection. Because transient interventions that directly alter paradigm adoption break this invariant, they shift the system to a different level set, producing persistent changes in the long-run cultural composition. The model thus reveals how societies retain a deterministic memory of prior health and treatment states, demonstrating that both misinformation shocks and targeted public health interventions can leave lasting structural changes long after the perturbation has ended.
\end{abstract}

\section{Introduction}
\label{sec:intro}

Disease dynamics do not unfold independently of human behavior. Individuals respond to perceived risk, observed outcomes, institutional recommendations, and the behavior of others, while these responses in turn modify subsequent health trajectories. This reciprocal dependence has motivated a broad family of coupled behavior--disease models aimed at understanding how individual decisions reshape population-level epidemiological outcomes \cite{funk2010modelling, fenichel2011adaptive, verelst2016behavioural, perra2021non}.

Most of this literature has been developed in the context of infectious diseases, where behavior modifies exposure, contact patterns, vaccination, or protection against a pathogen transmitted through the population. In vaccination games, this interaction may additionally generate a free-rider dilemma: individuals benefit from the protection produced by others while avoiding the private cost or perceived risk of preventive action \cite{bauch2004vaccination, wang2016statistical}. These mechanisms do not transfer directly to recurrent or non-communicable conditions, where incidence may be exogenous or independent across individuals and no collective immunity is generated. Nevertheless, the absence of biological transmission does not imply the absence of social transmission: beliefs about treatments, their expected benefits, and their risks may still spread through interpersonal influence and social reinforcement \cite{goldman2015social, lattenaor2018influence}.

Treatment preferences are therefore not determined by comparative clinical effectiveness alone. They may also be shaped by trust, prior beliefs, personal experience, expectations, social identity, and the endorsement of relatives or other members of an individual's social environment \cite{astin1998why,goldman2015social,lattenaor2018influence}. Clinical effectiveness and cultural attractiveness are consequently distinct quantities. A treatment may provide poorer average health outcomes while remaining socially visible and culturally competitive, particularly when anecdotal successes are more readily communicated than unsuccessful or uneventful experiences.

Alternative medicine offers no shortage of candidate case studies for this separation; homeopathy is paradigmatic among them---though by no means the only instance---precisely because of the scale and persistence of its acceptance: surveys across eleven countries report that a non-negligible fraction of the population uses homeopathy each year, with rates as high as 9.8\% where public health systems provide coverage \cite{relton2017prevalence}. Broad evidence assessments have not established reliable efficacy for specific health conditions, while systematic reviews of randomized placebo-controlled trials have reported a small effect for individualized treatment and no clear effect for non-individualized treatment \cite{nhmrc2015evidence,mathie2014randomised,mathie2017randomised}. However, subsequent critical evaluations have emphasized that these signals are highly sensitive to methodological limitations and selective reporting, with substantial reporting bias affecting the available trial literature \cite{emprechtinger2022assessing}. This fragility is not merely theoretical: an influential meta-analysis restricted its central estimate to eight trials drawn from an initial pool of 110 and reported effects consistent with placebo \cite{shang2005are}, yet subsequent sensitivity analyses showed that the statistical conclusion was highly dependent on the choice of trial subset, with alternative defensible subsets yielding statistically significant effects \cite{ludtke2008conclusions}. The clinical evidence base for homeopathy is thus not merely weak but unstable under scrutiny \cite{ernst2002systematic}. The relevant public-health concern is not complementary use in itself, but the possibility that confidence in an unsupported or less effective treatment substitutes for effective care. Among patients with curable cancers, complementary-medicine use was associated with poorer survival in an association that was no longer statistically significant after accounting for treatment refusal or delay \cite{johnson2018complementary}. The competition between treatment paradigms may therefore have genuine health consequences whenever cultural adoption alters which treatment individuals ultimately receive.

This raises a basic epidemiological question. Consider a population in which individuals experiencing the same recurrent condition divide between two treatment paradigms with different recovery effectiveness, while their adoption is also shaped by social influence. Does the population necessarily converge toward the clinically superior paradigm? Or can cultural transmission sustain the coexistence, or even the predominance, of a less effective alternative? If several long-run cultural compositions are possible, what determines which one is selected, and can temporary changes in health conditions or social influence leave persistent effects? Addressing these questions requires treating disease burden and paradigm adherence not as separate quantities---one biological and one cultural---but as components of a single coupled dynamical system. In the minimal framework developed below, we close this feedback loop by assuming that the social attractiveness of each paradigm responds to changes in its associated illness burden. We additionally allow health outcomes associated with the two paradigms to differ in their social visibility, represented by paradigm-specific amplification coefficients.

In brief, we introduce a deterministic mean-field model coupling a recurrent non-infectious condition to the cultural competition between two treatment paradigms. Treatment-specific recovery rates govern the health dynamics, whereas changes in the corresponding illness burdens shape cultural attractiveness through paradigm-specific social amplification coefficients. We show that the resulting system possesses a one-dimensional transversely attracting manifold of equilibria and, for interior cultural compositions, an exact conserved quantity. The health variables relax towards stationary burdens, while the final cultural composition is selected by the initial cultural and health state. Asymmetry in social amplification modifies this selection but is not required for the equilibrium manifold to exist. Transient perturbations that break the invariant can consequently displace the population from one compatible equilibrium to another, leaving persistent changes in the ideal neutral model.

The remainder of the paper is organized as follows. Section~\ref{sec:model} introduces the coupled biological and cultural dynamics and discusses their main assumptions. Section~\ref{sec:analytical} derives the equilibrium manifold, its transverse stability, and the conserved quantity governing equilibrium selection. We then examine the transient dynamics, path-dependent responses to perturbations, and parameter sensitivity in Section~\ref{sec:results}. Finally, Section~\ref{sec:discussion} discusses the interpretation, limitations, and possible extensions of the framework.

\section{The model}
\label{sec:model}
We formulate a deterministic mean-field model to describe the interplay between individual health states and cultural paradigms.

\subsection{Assumptions}
The model is built upon the following core assumptions:
\begin{enumerate}
    \item \textbf{Non-Infectious Disease:} The transition from a healthy state to an ill state is independent of the number of currently ill individuals. There is no biological contagion.
    \item \textbf{Health States:} Individuals are strictly classified as either healthy ($H$) or ill ($I$). 
    \item \textbf{Treatment choices:} Each individual is associated with one of two treatment approaches, evidence-based medicine ($M$) or alternative medicine ($A$). This association represents the treatment the individual would adopt when affected by the focal condition and may change through social influence. The corresponding population shares satisfy $M+A=1$.
    \item \textbf{Recovery Rates:} The recovery rate is strictly dependent on the adopted paradigm, with evidence-based medicine providing a faster recovery: $\gamma_M > \gamma_A$.
    \item \textbf{Social-amplification asymmetry}: Health outcomes associated with the two treatments may differ in their social amplification, represented by non-negative coefficients $r_M$ and $r_A$. We focus primarily on the asymmetric case $r_A>r_M$, while noting that this ordering is not required for the equilibrium manifold to exist.
\end{enumerate}

\subsubsection*{Mean-field description of treatment switching}
The behavioral variable $M$ describes the aggregate population share associated with evidence-based medicine, with $A=1-M$ denoting the
corresponding alternative-medicine share. Changes in these quantities are treated at the population level through the replicator equation rather than through explicit individual transfers between health--treatment compartments. Accordingly, the reduced model does not resolve the health status of the particular individuals whose treatment association changes during the behavioral dynamics.

The variables $I_M$ and $I_A$ should therefore be interpreted as the aggregate illness burdens associated with the two treatment groups, whose evolution is determined by incidence and treatment-specific recovery for the instantaneous population shares $M$ and $A$. The accompanying healthy fractions are defined by $H_M=M-I_M$ and $H_A=A-I_A$. In this reduced mean-field closure, changes in treatment composition modify the sizes of the populations exposed to each treatment-specific health dynamics, while the microscopic redistribution of healthy and ill individuals during individual switching events is not represented explicitly.

A fully disaggregated description could instead introduce separate healthy and ill switching fluxes between the two treatment groups, allowing switching propensities to depend explicitly on current health status. Such a formulation would require additional assumptions and state-transition terms and need not preserve the exact dynamical structure studied here. Since our present objective is to analyze the minimal feedback between aggregate treatment choice and treatment-dependent illness burden, we retain the reduced three-dimensional formulation and revisit this limitation in Sec.~\ref{sec:discussion}.

\subsubsection*{Constant population and time-scale separation}
Our framework intentionally omits demographic processes, such as natural births and disease-induced or natural mortality, and normalizes the total population to unity ($N=1$). This abstraction can be motivated through a separation of time scales. The primary dynamics of interest---the social transmission of treatment preferences and the clinical recovery from the non-infectious condition---may unfold over timescales considerably shorter than those associated with substantial demographic turnover. The present model is therefore intended to isolate this faster coupled health--behavior dynamics before introducing slower population-renewal processes.

Demographic turnover would constitute an additional source of forcing by continually introducing individuals whose treatment preferences and health states need not follow the invariant structure of the closed population. Mathematically, such processes would generally break the exact conservation law derived below and therefore remove the perfect neutrality of the idealized system. Depending on how births, deaths, and preference formation are modeled, this neutrality breaking could generate slow drift along the former equilibrium manifold, select particular long-run compositions, or replace the continuous family of equilibria by a smaller set of attractors. The precise outcome cannot be determined without specifying the demographic mechanism explicitly.

We therefore neglect demographic turnover not because it is irrelevant in real populations, but because our objective is to expose the endogenous mechanism generated by the coupling between illness dynamics and treatment competition in its minimal form.

\subsection{Biological dynamics}
Let $H_k$ and $I_k$ denote the fraction of healthy and ill individuals within paradigm $k \in \{M, A\}$. Assuming a constant per capita incidence rate $\beta$, the biological transitions are governed by linear ordinary differential equations:

\begin{align}
    \frac{dI_M}{dt} &= \beta H_M - \gamma_M I_M \label{eq:ill_M}  \, ,\\
    \frac{dI_A}{dt} &= \beta H_A - \gamma_A I_A \label{eq:ill_A} \, .
\end{align}

Since $H_k + I_k = k$ for a given paradigm size, the dynamics of the healthy populations are implicitly defined as $\dot{H}_k = \dot{k} - \dot{I}_k$.

\subsection{Behavioral replicator dynamics}

Standard epidemiological models (e.g., SIS, SIR) typically frame human behavior as a direct modifier of the transmission rate $\beta$ within a biological contagion process. In our framework, because the disease is non-infectious, the contagion occurs strictly in the space of beliefs, i.e., treatment choices.

To model this transmission, we utilize a replicator equation. Rather than imposing a standard game-theoretic payoff matrix, we define the rate of adoption of a paradigm based on its social \textit{attractiveness}, or convincing power, denoted by $\alpha_k$ ($k=M,A$). The shift toward a strategy $M$ against an alternative $A$ is driven by the difference in their respective attractiveness:
\begin{equation}
\frac{dM}{dt} = M A [\alpha_M - \alpha_A] \, .
\end{equation}

In this context, the social attractiveness of a treatment for the background disease is driven by visible changes in its associated health outcomes. We assume that improvements or deteriorations in the illness burden contribute to treatment attractiveness through a treatment-specific social-amplification coefficient $r_k\geq 0$. A declining illness burden generates a positive social signal, whereas an increasing burden generates a negative one, with $r_k$ controlling the strength with which these changes are socially amplified. We therefore define $\alpha_M \equiv -r_M\dot{I}_M$ and $\alpha_A \equiv -r_A\dot{I}_A$.

Substituting these terms into the replicator equation clarifies the mathematical origin of the signs. By explicitly writing the double negatives, we obtain:

\begin{equation}
\frac{dM}{dt} = M A \left[ \left( - r_M \frac{dI_M}{dt} \right) - \left( - r_A \frac{dI_A}{dt} \right) \right] \, .
\end{equation}

Resolving the signs and recalling that $A = 1 - M$, we yield our governing equation for cultural transmission:
\begin{equation}
\frac{dM}{dt} = M (1 - M) \left[ r_A \frac{dI_A}{dt} - r_M \frac{dI_M}{dt} \right]
\label{eq:replicator} \, .
\end{equation}

This derivation also clarifies what may initially seem counter-intuitive: the drive toward evidence-based medicine ($M$) depends on the relative social signals generated by changes in the illness burdens associated with the two treatments, rather than directly on their absolute illness levels.

Crucially, grounding the social attractiveness exclusively in the derivative of the disease burden (rather than the absolute stock of healthy or ill individuals) is motivated by psychological and sociological observations related to \textit{event-driven salience}. In health-related decision-making, cognitive biases such as the availability heuristic \cite{tversky1973availability} suggest that salient events may disproportionately influence individual judgments. Instead, social discourse and word-of-mouth advocacy may be particularly responsive to salient, transient events---specifically, sudden recoveries or unexpected illnesses \cite{berger2011arousal}. Once a health state becomes chronic, it loses some of its conversational novelty; the individual habituates, and their active advocacy may diminish. Therefore, within our modeling framework, a medical paradigm only gains social traction when it generates an active, observable epidemiological momentum. This derivative-based formulation provides a stylized macroscopic representation of event-sensitive social transmission, in which information contagion is driven by dynamic changes in state rather than by the static states themselves \cite{centola2010spread}.

Furthermore, our formulation evaluates this epidemiological momentum at the macroscopic aggregate level ($\alpha_k = - r_k \dot{I}_k$) rather than as a per-capita rate (e.g., $- r_k \dot{I}_k / k$), as is customary in standard evolutionary game theory. This departure is intentional and provides a stylized representation of the aggregate mechanics of social transmission. The cultural visibility of a medical paradigm is driven by the absolute volume of positive anecdotes and testimonies circulating in the public sphere, rather than by its true statistical probability of success. Due to the availability heuristic, individuals are more likely to be persuaded by a dominant paradigm that produces a large total number of visible events (even if its per-capita efficacy is mediocre) than by a niche paradigm with a high per-capita success rate but too few total adopters to generate a noticeable volume of social signals. Consequently, aggregate momentum represents the total persuasive weight that each paradigm is assumed to exert on the population. See Appendix \ref{appSection:alternativeModel} for a further discussion of how tracking dynamic fluxes rather than absolute health states changes the equilibrium structure.

\subsection{Coupled socio-epidemiological dynamics}
The integration of the biological equations (Eq. \ref{eq:ill_M} and \ref{eq:ill_A}) with the cultural equation (Eq. \ref{eq:replicator}) generates a fully coupled socio-epidemiological system. Rather than human behavior acting as a static parameter, the coupling establishes a continuous, bidirectional feedback loop.
    
First, the macroscopic cultural state (the distribution of $M$ and $A$) determines the effective recovery rate of the global population, as the population shares associated with the two paradigms dictate how many individuals are subject to the faster recovery rate $\gamma_M$ versus the slower rate $\gamma_A$.

Second, the resulting biological dynamics (the differential growth rates of $I_M$ and $I_A$) feed directly back into the cultural layer. The macroscopic health outcomes dynamically update the payoff differential in the replicator equation, driving further shifts in the population's paradigm distribution. This coupling implies that, although the biological dynamics are linear for a fixed treatment composition, coupling them to treatment choice through the replicator equation renders the full system nonlinear.

\section{Equilibrium and stability analysis}
\label{sec:analytical}

To analyze the stability of the coupled socio-epidemiological system, we must define the full state vector and derive its Jacobian matrix. We express the healthy populations in terms of the adopted paradigms and the illness burdens: $H_M = M - I_M$ and $H_A = 1 - M - I_A$.

The state vector is defined as $\vec{X} = (M, I_M, I_A)^T$, and the complete autonomous system is governed by:
\begin{align}
    f_1 (\vec{X}) &= \dot{M} = M(1 - M) \left[ r_A \dot{I}_A - r_M \dot{I}_M \right] \label{eq:sys_M} \, , \\
    f_2 (\vec{X}) &= \dot{I}_M = \beta (M - I_M) - \gamma_M I_M \label{eq:sys_IM} \, , \\
    f_3 (\vec{X}) &= \dot{I}_A = \beta (1 - M - I_A) - \gamma_A I_A \label{eq:sys_IA} \, .
\end{align}

\subsection{Continuous manifolds of equilibria}
An equilibrium point $\vec{X}^* = (M^*, I_M^*, I_A^*)$ requires $\dot{M} = 0$, $\dot{I}_M = 0$, and $\dot{I}_A = 0$. 

Setting equations \ref{eq:sys_IM} and \ref{eq:sys_IA} to zero yields the steady-state biological burdens for any given macroscopic cultural distribution $M^*$:
\begin{align}
    I_M^* &= \left( \frac{\beta}{\beta + \gamma_M} \right) M^* \label{eq:eq_IM}  \, ,\\
    I_A^* &= \left( \frac{\beta}{\beta + \gamma_A} \right) (1 - M^*) \label{eq:eq_IA} \, .
\end{align}

Substituting the condition that $\dot{I}_M = 0$ and $\dot{I}_A = 0$ into equation \ref{eq:sys_M}, we find that $\dot{M} = 0$ is automatically satisfied for any value of $M^* \in [0, 1]$. This implies that the system does not possess isolated equilibrium points. The equilibria therefore form a continuous, one-dimensional manifold (a line segment) parameterized by $M^*$. The dynamical selection of a particular point on this manifold is addressed below.

\subsection{The Jacobian matrix construction}
To evaluate local stability, we construct the Jacobian matrix $J_{ij} = \partial f_i / \partial X_j$ evaluated at $\vec{X}^*$.\\

We first calculate the partial derivatives of the biological subsystem (the second and third rows of $J$):
\begin{equation}
\begin{aligned}
    J_{21} &= \frac{\partial f_2}{\partial M} = \beta\, , \quad &J_{22} &= \frac{\partial f_2}{\partial I_M} = -(\beta + \gamma_M)\, , \quad &J_{23} &= \frac{\partial f_2}{\partial I_A} = 0 \, , \\
    J_{31} &= \frac{\partial f_3}{\partial M} = -\beta\, , \quad &J_{32} &= \frac{\partial f_3}{\partial I_M} = 0\, , \quad &J_{33} &= \frac{\partial f_3}{\partial I_A} = -(\beta + \gamma_A) \, .
\end{aligned}
\end{equation}

For the first row (the cultural dynamics), applying the product rule to $f_1$ yields:
\begin{equation}
    \frac{\partial f_1}{\partial X_j} = \frac{\partial [M(1 - M)]}{\partial X_j} \left[ r_A f_3 - r_M f_2 \right] + M(1 - M) \left[ r_A \frac{\partial f_3}{\partial X_j} - r_M \frac{\partial f_2}{\partial X_j} \right] \, .
\end{equation}

Crucially, at any equilibrium point $\vec{X}^*$, $f_2 = 0$ and $f_3 = 0$. Therefore, the first term vanishes entirely. Recognizing that $\partial f_2 / \partial X_j = J_{2j}$ and $\partial f_3 / \partial X_j = J_{3j}$, the components of the first row simplify precisely to:
\begin{equation}
    J_{1j} = M^*(1 - M^*) \left[ r_A J_{3j} - r_M J_{2j} \right] \quad \text{for } j \in \{1, 2, 3\} \, .
\end{equation}

\subsection{Non-hyperbolicity}
The algebraic structure derived above reveals a fundamental topological property of the model: at equilibrium, the entire first row of the Jacobian matrix is a strict linear combination of the second and third rows.

Consequently, the rows of $J(\vec{X}^*)$ are linearly dependent, which mathematically dictates that the determinant is zero:
\begin{equation}
    \det(J) = 0 \, ,
\end{equation}
which confirms the existence of at least one zero eigenvalue ($\lambda_1 = 0$).

The presence of a zero eigenvalue implies that every equilibrium is non-hyperbolic. Accordingly, the Hartman--Grobman theorem is not applicable at these equilibria. The zero eigenvalue corresponds to the neutral stability along the continuous manifold of cultural equilibria ($M^*$), proving analytically that while the biological incidence transversally attracts the system to a fixed disease burden (the remaining eigenvalues $\lambda_2, \lambda_3$ have negative real parts), the overarching paradigm distribution remains fundamentally neutral and socially negotiated.

\subsection{Constant of motion and equilibrium selection}
\label{sec:constant_motion}
The continuous equilibrium manifold can be further characterized through an exact constant of motion. For an interior treatment share, $0<M<1$, the replicator equation~\eqref{eq:replicator} can be written as
\begin{equation}
    \frac{1}{M(1-M)}\frac{dM}{dt}
    =
    r_A\frac{dI_A}{dt}
    -
    r_M\frac{dI_M}{dt}\, .
    \label{eq:replicator_separable}
\end{equation}
Using partial fraction decomposition, the left-hand side can be expressed as
\begin{equation}
    \frac{d}{dt}
    \ln\left(\frac{M}{1-M}\right)
    =
    r_A\dot I_A-r_M\dot I_M\, .
\end{equation}
Hence,
\begin{equation}
    \frac{d}{dt}
    \left[
        \ln\left(\frac{M}{1-M}\right)
        +r_M I_M-r_A I_A
    \right]
    =0\, .
\end{equation}
The quantity
\begin{equation}
    \mathcal{C}
    =
    \ln\left(\frac{M}{1-M}\right)
    +r_M I_M-r_A I_A
    \label{eq:conserved_quantity}
\end{equation}
is therefore conserved along every trajectory contained in the interior region $0<M<1$. This algebraic invariant constrains the dynamics geometrically: the interior state space is partitioned into a continuous family of two-dimensional invariant level sets, each identified by a value of $\mathcal{C}$.

For an initial state $\mathbf{X}(0)=(M_0,I_{M,0},I_{A,0})$, the corresponding value of the constant is
\begin{equation}
    \mathcal{C}_0
    =
    \ln\left(\frac{M_0}{1-M_0}\right)
    +r_M I_{M,0}
    -r_A I_{A,0}\, .
    \label{eq:initial_constant}
\end{equation}
Consequently, each interior trajectory is confined to a two-dimensional level set $\mathcal{C}=\mathcal{C}_0$ in the three-dimensional state space.

The conserved quantity also provides an explicit selection rule for the equilibrium compatible with a given initial condition. Defining
\begin{equation}
    a_M\equiv\frac{\beta}{\beta+\gamma_M},
    \qquad
    a_A\equiv\frac{\beta}{\beta+\gamma_A}\, ,
\end{equation}
the equilibrium manifold derived in Eqs.~\eqref{eq:eq_IM}--\eqref{eq:eq_IA} is
\begin{equation}
    I_M^*=a_M M^*,
    \qquad
    I_A^*=a_A(1-M^*)\, .
\end{equation}
Substitution into Eq.~\eqref{eq:conserved_quantity} gives the value of the constant along the equilibrium manifold,
\begin{equation}
    \Phi(M^*)
    =
    \ln\left(\frac{M^*}{1-M^*}\right)
    +r_M a_M M^*
    -r_A a_A(1-M^*)\, .
    \label{eq:equilibrium_selection_function}
\end{equation}
Thus, an equilibrium belonging to the same invariant level set as the initial state must satisfy
\begin{equation}
    \Phi(M^*)=\mathcal{C}_0\, .
    \label{eq:equilibrium_selection}
\end{equation}

This compatible equilibrium is unique. Indeed,
\begin{equation}
    \frac{d\Phi}{dM}
    =
    \frac{1}{M(1-M)}
    +r_M a_M+r_A a_A
    >0\, ,
    \qquad 0<M<1\, ,
    \label{eq:phi_monotonic}
\end{equation}
for $r_M,r_A\geq0$. Moreover,
\begin{align}
    \Phi(M)\rightarrow-\infty
    \quad\text{as}\quad M\rightarrow0^+\, ,\\
    \qquad
    \Phi(M)\rightarrow+\infty
    \quad\text{as}\quad M\rightarrow1^-\, .
\end{align}
Thus, $\Phi:(0, 1)\rightarrow\mathbb{R}$ is a strictly increasing bijection. Consequently, every interior initial condition defines exactly one equilibrium on the same invariant level set. If the corresponding trajectory converges to the equilibrium manifold, this point is necessarily its asymptotic state, with $M_\infty$ determined by Eq.~\eqref{eq:equilibrium_selection}.

The continuous manifold therefore represents the set of possible equilibria, whereas the conserved quantity determines which point on that manifold is compatible with a given initial state. The equilibrium selection depends on the complete initial state $(M_0,I_{M,0},I_{A,0})$, rather than on the initial treatment share alone, as well as on the social-amplification coefficients $r_M$ and $r_A$. In particular, asymmetric social amplification is not required for this selection mechanism: the conserved quantity and the equilibrium manifold remain well defined when $r_M=r_A$.

A transient external perturbation that drives the system away from its unforced trajectory may change the value of $\mathcal{C}$ and thereby place the system on a different invariant level set, associated with a different equilibrium.

The boundary sets $M=0$ and $M=1$ are invariant but are not described by Eq.~\eqref{eq:conserved_quantity}, since the logarithmic term diverges there. They must therefore be considered separately from the interior dynamics.

\section{Results}
\label{sec:results}

\subsection{Transient approach to the equilibrium manifold}
\label{sec:transient}

\begin{figure*}[htpb]
    \centering
    \includegraphics[width=\linewidth]{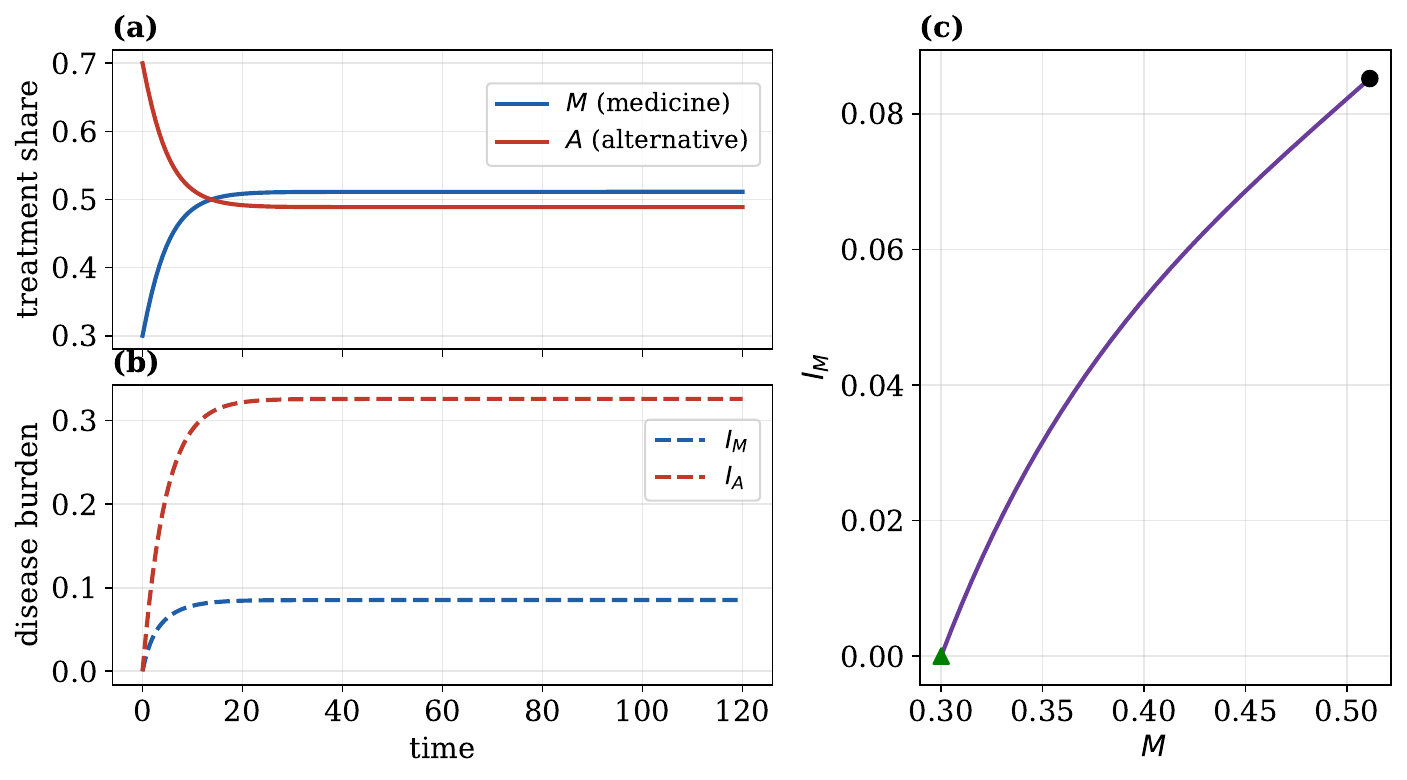}
    \caption{\textbf{Transient dynamics from an initially healthy population.} \textbf{(a)} Treatment shares $M(t)$ and $A(t)$; \textbf{(b)} illness burdens $I_M(t)$ and $I_A(t)$, sharing the time axis with panel (a); \textbf{(c)} phase-space projection $M$ vs.\ $I_M$. For this initial condition and parameter set, all displayed variables approach their asymptotic values monotonically. The local approach to the equilibrium manifold is non-oscillatory, consistently with the real and negative transverse eigenvalues derived in Sec.~\ref{sec:analytical}. Parameters used are $\beta=0.1$, $\gamma_M=0.5$, $\gamma_A=0.05$, $r_M=1$, and $r_A=3$.}
    \label{fig:transient}
\end{figure*}

Figure~\ref{fig:transient} illustrates the relaxation of the coupled system from an out-of-equilibrium initial condition. As shown in panel~(a), the treatment shares $M(t)$ and $A(t)$ evolve during the transient, while the illness burdens in panel~(b) approach the stationary values $I_M^*$ and $I_A^*$ given by Eqs.~\eqref{eq:eq_IM}--\eqref{eq:eq_IA}. The phase-space projection in panel~(c) shows the trajectory approaching a point on the continuous equilibrium manifold, the numerical signature of the neutral manifold derived above. The final treatment share is not selected by an isolated attractor, but by the invariant level set associated with the initial state.

The numerical trajectory shown in Fig.~\ref{fig:transient} therefore illustrates the analytical structure derived in the previous section. The equilibrium manifold specifies the continuum of stationary states available to the system, whereas the conserved quantity $\mathcal{C}$ determines which point on that manifold is compatible with a particular initial condition. In other words, the final treatment share is not an arbitrary value acquired when the biological transient happens to cease. For an interior trajectory, it is constrained from the outset by
\begin{equation}
    \Phi(M_\infty)=\mathcal{C}_0\, ,
\end{equation}
provided that the trajectory converges to the equilibrium manifold.

For the trajectory displayed here, both the treatment shares and illness burdens evolve monotonically. This monotonicity of the individual state variables is a property of the particular trajectory and should be distinguished from the more general local stability result. What follows analytically from the Jacobian spectrum is that the equilibrium manifold is transversely attracting and that the two non-zero eigenvalues are strictly real and negative. Thus, sufficiently close to the manifold, perturbations decay without a rotational or spiraling component in the transverse directions. 

This absence of oscillation is a structural guarantee rather than an artifact of the chosen parameters. To see this explicitly, let $p \equiv M^*(1-M^*)$ and recall that, since $\lambda_1 = 0$, the remaining eigenvalues satisfy $\lambda_2+\lambda_3 = \mathrm{tr}(J)$ and $\lambda_2\lambda_3 = S$, where $S$ is the sum of the three principal $2\times 2$ minors of $J$. A direct computation from the Jacobian entries in Sec.~\ref{sec:analytical} gives
\begin{equation}
    \mathrm{tr}(J) = -\big[p\beta(r_A+r_M) + (\beta+\gamma_M) + (\beta+\gamma_A)\big]\, ,
\end{equation}
\begin{equation}
    S = (\beta+\gamma_M)(\beta+\gamma_A) + p\beta\big[r_M(\beta+\gamma_A) + r_A(\beta+\gamma_M)\big]\, .
\end{equation}
The transverse eigenvalues are therefore the roots of $\lambda^2 - \mathrm{tr}(J)\,\lambda + S = 0$, with discriminant $D = \mathrm{tr}(J)^2 - 4S$. Writing $w = \gamma_M - \gamma_A$ and $X = p\beta(r_A+r_M)$, algebraic simplification of $D$ yields
\begin{equation}
    D = w^2 - 2p\beta(r_A - r_M)\,w + X^2\, ,
    \label{eq:discriminant_quadratic}
\end{equation}
a quadratic expression in $w$ with unit leading coefficient. Its own discriminant, taken with respect to $w$, is
\begin{equation}
    D_w = \big[2p\beta(r_A-r_M)\big]^2 - 4X^2 = -16\,(p\beta)^2\, r_A r_M\, .
\end{equation}

Since $r_M, r_A >0$ in the parameter regime considered here, together with $\beta>0$ and $p=M^*(1-M^*)>0$ for any interior equilibrium, we have $D_w<0$ strictly. Because the quadratic \eqref{eq:discriminant_quadratic} has a positive leading coefficient and a negative discriminant in $w$, it has no real roots and therefore remains strictly positive. Hence $D>0$, and the transverse eigenvalues $\lambda_2$, $\lambda_3$ are guaranteed to be real throughout the non-degenerate parameter regime $r_M ,r_A>0$, not only for the parameter set used in Fig.~\ref{fig:transient}. If the degenerate cases $r_M=0$ or $r_A =0$ are also allowed, $D$ remains non-negative, although it may vanish for isolated parameter combinations.


\subsection{Equilibrium selection and persistent path dependence}
\label{sec:variety}

\begin{figure*}[htbp]
    \centering
    \includegraphics[width=\linewidth]{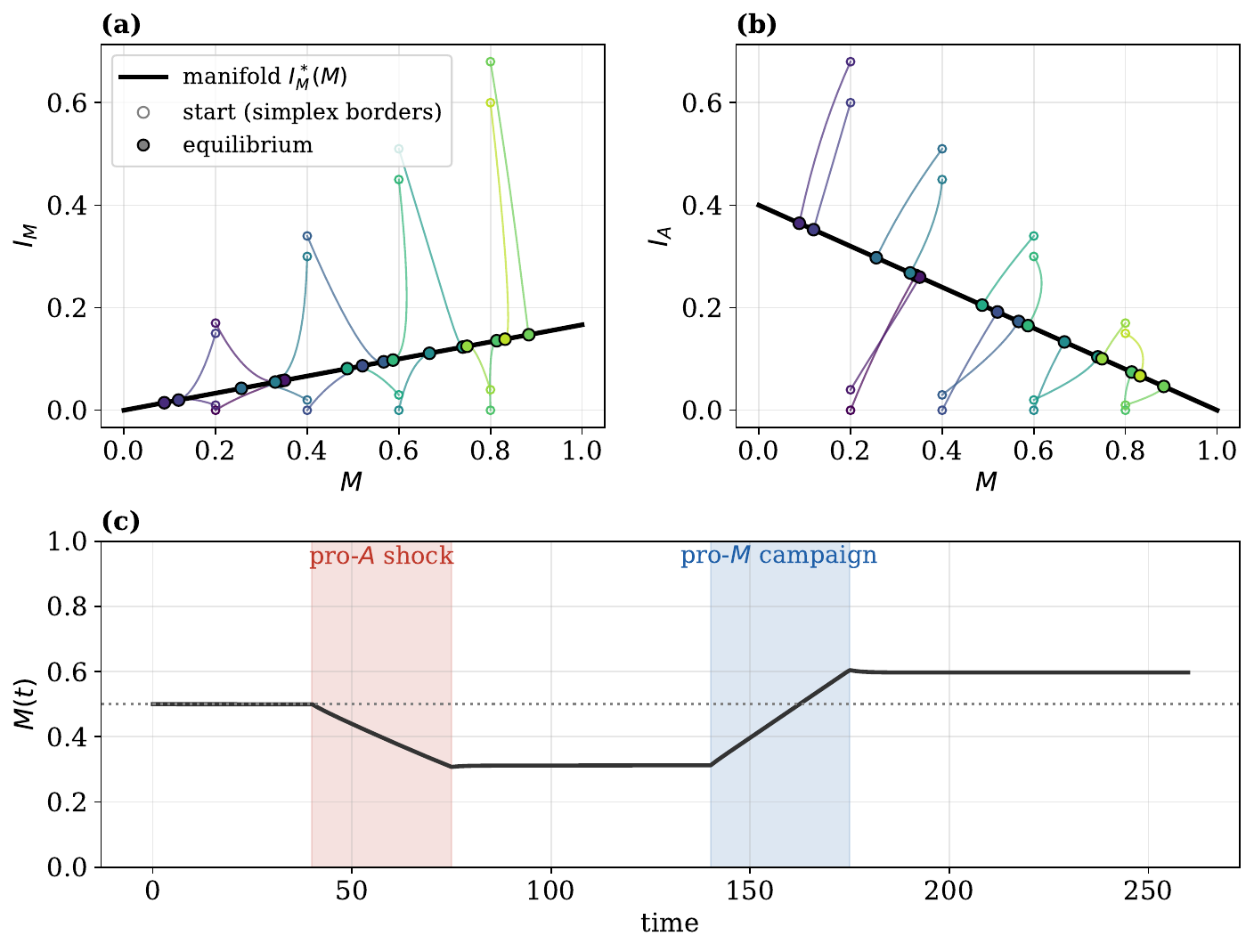}
    \caption{\textbf{Continuous manifold of equilibria and path-dependent equilibrium selection.} Panels \textbf{(a)} and \textbf{(b)}: two projections of the phase space, $M$ vs.\ $I_M$ and $M$ vs.\ $I_A$, respectively. The thick black line is the one-dimensional equilibrium manifold obtained from Eqs.~\eqref{eq:eq_IM}--\eqref{eq:eq_IA}. Thin colored curves show RK4 trajectories starting from different initial states (open circles), each approaching the unique equilibrium compatible with its invariant level set (solid circles). Apparent intersections between trajectories arise solely from projection onto two dimensions; trajectories remain distinct in the full three-dimensional state space. Panel \textbf{(c)}: persistent displacement produced by direct external interventions on treatment adoption. Starting from an established equilibrium, a transient pro-$A$ intervention decreases $M$ (red band), whereas a subsequent pro-$M$ intervention increases it (blue band). Because the intervention acts directly on the treatment share, it breaks conservation of $\mathcal{C}$ while active, placing the system on a different invariant level set. Once the forcing is removed, the autonomous dynamics relax toward the equilibrium associated with the new value of $\mathcal{C}$. Parameters are $\beta=0.1$, $\gamma_M=0.5$, $\gamma_A=0.15$, $r_M=1$, and $r_A=3$.}
    \label{fig:variety_hysteresis}
\end{figure*}

Figure~\ref{fig:variety_hysteresis} makes the geometry of equilibrium selection explicit. Panels (a) and (b) show two projections of the same three-dimensional phase space. The solid black lines correspond to the analytical equilibrium manifold,
\begin{equation}
    I_M^*=a_M M^*\, ,
    \qquad
    I_A^*=a_A(1-M^*)\, ,
\end{equation}
whereas the colored curves correspond to trajectories launched from different initial states. Although the trajectories shown approach the same one-dimensional family of equilibria, distinct initial states generally select distinct points on that family.

The conserved quantity derived in Sec.~\ref{sec:constant_motion} provides the precise mathematical interpretation of this observation. Each interior initial condition defines a value $\mathcal{C}_0$ and therefore confines the trajectory to the corresponding two-dimensional invariant level set. Because the restriction of the conserved quantity to the equilibrium manifold, $\Phi(M)$, is strictly increasing, each such invariant surface intersects the equilibrium manifold at exactly one point. Consequently, the numerical endpoints shown in panels (a) and (b) are not arbitrary frozen states: they are the unique equilibria compatible with their respective initial conditions.

This also clarifies the meaning of the neutral direction associated with $\lambda_1=0$. Neutrality along the manifold does not imply that an unperturbed trajectory may move freely among its equilibria. Once the full initial state $(M_0,I_{M,0},I_{A,0})$ is specified, the invariant $\mathcal{C}_0$ restricts the dynamics to one particular level set and thereby selects one compatible equilibrium. The continuum arises because different initial states correspond to different values of $\mathcal{C}_0$, not because a single autonomous trajectory can wander arbitrarily along the manifold.

The same structure explains the persistent response to external interventions shown in panel~(c). Here the system is first allowed to settle on the equilibrium manifold and is subsequently subjected to a direct external forcing of the treatment share. Schematically, during such an intervention the behavioral equation becomes
\begin{equation}
    \dot M
    =
    M(1-M)
    \left(
        r_A\dot I_A-r_M\dot I_M
    \right)
    +u(t)\, ,
    \label{eq:forced_M}
\end{equation}
where $u(t)$ denotes the externally imposed shift in treatment adoption. The conserved quantity is then no longer constant. Instead,
\begin{equation}
    \dot{\mathcal{C}}
    =
    \frac{u(t)}{M(1-M)}\, .
    \label{eq:forced_C}
\end{equation}
The intervention therefore transfers the system from one invariant level set to another. Once $u(t)$ vanishes, conservation is restored, but now at the new value of $\mathcal{C}$. The subsequent autonomous dynamics must therefore approach a different compatible equilibrium.

Panel~(c) illustrates this mechanism through two successive interventions. A pro-alternative perturbation first reduces the evidence-based medicine share, whereas a later pro-medicine intervention shifts it in the opposite direction. After each intervention the system retains a persistent displacement relative to its preceding equilibrium, because the forcing has changed the invariant level set on which the unforced dynamics subsequently evolves. This provides a precise dynamical mechanism for history-dependent equilibrium selection.

The persistence shown here is closely related to hysteresis or remanent behavior in the broad sense that the state reached after a perturbation depends on the system's previous evolution. Strictly speaking, however, Fig.~\ref{fig:variety_hysteresis} does not implement a conventional closed hysteresis loop obtained by cyclically varying a control parameter. We therefore refer primarily to \emph{path dependence} or \emph{persistent displacement} in what follows.

An important consequence of the conserved quantity is that not every temporary perturbation necessarily produces such a permanent relocation. For example, a transient change in the incidence rate $\beta(t)$ does not by itself alter $\mathcal{C}$ provided that the derivative-coupled behavioral equation remains valid throughout the perturbation. Indeed, the derivation of Eq.~\eqref{eq:conserved_quantity} does not require $\beta$ to be constant: the terms involving $\dot I_M$ and $\dot I_A$ continue to cancel exactly. Such a perturbation can generate a substantial transient excursion but, after the original parameters are restored, the system remains on the same invariant level set and is therefore associated with the same compatible equilibrium. Persistent relocation requires a perturbation that changes $\mathcal{C}$, such as the direct adoption forcing represented in Eq.~\eqref{eq:forced_M}, or another external mechanism that explicitly breaks the invariant relation.

\subsection{Parameter dependence and equilibrium selection}
\label{sec:sensitivity}

\begin{figure*}[htbp]
    \centering
    \includegraphics[width=\linewidth]{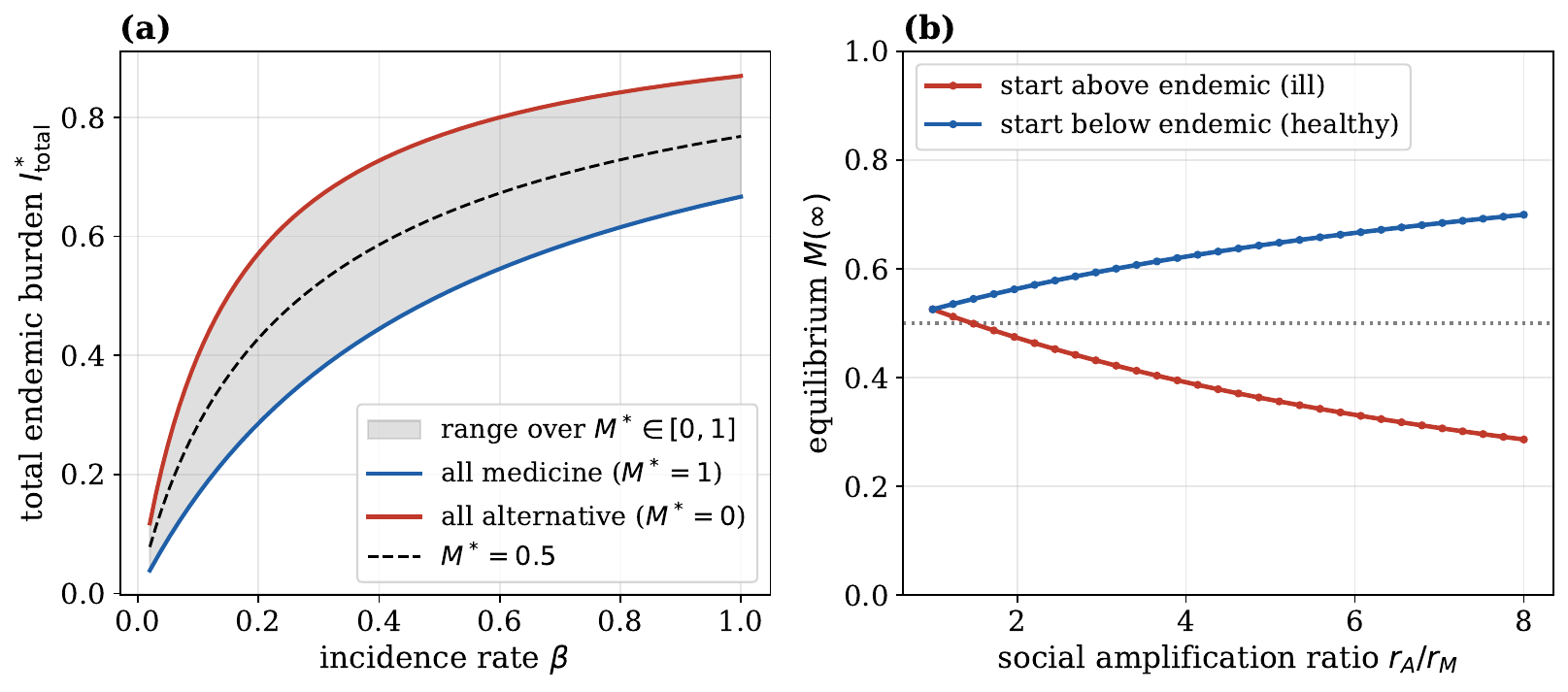}
    \caption{\textbf{Parameter dependence of the equilibrium burden and selected treatment share.} \textbf{(a)} Total equilibrium illness burden $I_{\mathrm{total}}^*=I_M^*+I_A^*$ as a function of the incidence rate $\beta$. The shaded region spans the complete equilibrium manifold, $M^*\in[0,1]$, and is bounded by the two limiting treatment compositions, $M^*=1$ (all medicine) and $M^*=0$ (all alternative treatment). \textbf{(b)} Selected equilibrium treatment share $M_\infty$ as a function of the social-amplification ratio $r_A/r_M$, obtained from two different out-of-equilibrium initial states. The blue curve corresponds to an initially non-ill population, whereas the red curve corresponds to a population with a high initial illness burden. The opposite dependence on $r_A/r_M$ reflects the fact that social amplification modifies the equilibrium-selection map through the complete initial state rather than shifting the equilibrium manifold itself. Fixed parameters are $\gamma_M=0.5$ and $\gamma_A=0.15$. Panel (b) additionally uses $\beta=0.1$ and $r_M=1$, with initial conditions $\mathbf{X}(0)=(0.5,0,0)$ and $\mathbf{X}(0)=(0.5,0.45,0.45)$.}
    \label{fig:sensitivity}
\end{figure*}

The analytical results allow two conceptually distinct forms of parameter dependence to be separated. Biological parameters such as the incidence rate $\beta$ and recovery rates $\gamma_M,\gamma_A$ determine the geometry and health burden of the equilibrium manifold itself. By contrast, the social-amplification coefficients $r_M$ and $r_A$ do not alter the existence or location of that manifold in $(M,I_M,I_A)$ space; rather, they modify the invariant surfaces and therefore the mapping between an initial state and the equilibrium selected on the manifold.

Figure~\ref{fig:sensitivity}(a) considers the first effect. Along the equilibrium manifold, the total illness burden is
\begin{equation}
    I_{\mathrm{total}}^*
    =
    I_M^*+I_A^*
    =
    a_M M^*
    +
    a_A(1-M^*).
    \label{eq:total_burden}
\end{equation}
For a fixed incidence rate $\beta$, the model therefore does not associate a single equilibrium illness burden with the population. Instead, the continuum of possible treatment compositions generates the interval shown by the shaded region in Fig.~\ref{fig:sensitivity}(a).

Because $\gamma_M>\gamma_A$, it follows that
\begin{equation}
    a_M<a_A
\end{equation}
and hence
\begin{equation}
    \frac{\partial I_{\mathrm{total}}^*}{\partial M^*}
    =
    a_M-a_A
    <0\, .
\end{equation}
Thus, within the assumptions of the model, equilibria with a larger evidence-based medicine share are associated with a smaller total illness burden. The upper and lower boundaries of the shaded region correspond respectively to the all-alternative and all-medicine limiting compositions. The shaded region should therefore be interpreted as the family of burdens available across the equilibrium manifold, rather than as uncertainty around a unique equilibrium prediction.

Figure~\ref{fig:sensitivity}(b) addresses a different question: how the social-amplification asymmetry changes the equilibrium selected from a fixed initial state. The numerical sweep varies $r_A/r_M$ while keeping $r_M$ fixed. Importantly, varying this ratio does not create, destroy, or displace the equilibrium manifold. It changes the invariant $\mathcal{C}_0$ and the function $\Phi$, thereby altering the particular intersection between the initial invariant surface and the manifold.

This dependence can also be obtained directly from the analytical selection condition. Let
\begin{equation}
    q\equiv\frac{r_A}{r_M}\, ,
\end{equation}
with $r_M$ fixed. Differentiating
$\Phi(M_\infty)=\mathcal{C}_0$ implicitly with respect to $q$ gives
\begin{equation}
    \frac{dM_\infty}{dq}
    =
    \frac{
        r_M
        \left[
            a_A(1-M_\infty)-I_{A,0}
        \right]
    }{
        \Phi'(M_\infty)
    }\, .
    \label{eq:sensitivity_rA}
\end{equation}
Since $\Phi'(M_\infty)>0$, the sign of the response is determined by the difference between the initial alternative-associated illness burden $I_{A,0}$ and its value at the selected equilibrium, $a_A(1-M_\infty)$.

This relation explains the opposite trends displayed in Fig.~\ref{fig:sensitivity}(b). For the initially non-ill population, $I_{A,0}=0$, and therefore
\begin{equation}
    \frac{dM_\infty}{dq}>0\, .
\end{equation}
Increasing the social amplification associated with the alternative treatment consequently shifts the selected equilibrium toward a larger evidence-based medicine share. In contrast, for the highly ill initial condition used in the figure, $I_{A,0}=0.45$, which exceeds the corresponding equilibrium burden throughout the parameter range shown. Equation~\eqref{eq:sensitivity_rA} therefore predicts
\begin{equation}
    \frac{dM_\infty}{dq}<0\, ,
\end{equation}
and the selected evidence-based medicine share decreases as the alternative amplification coefficient increases.

The numerical curves can thus be understood without invoking a bifurcation or a change in the stability structure of the model. They represent two different responses of the same continuous equilibrium manifold to changes in the equilibrium-selection map. The direction of the response depends on the complete initial biological and treatment state, consistently with the constant-of-motion analysis of Sec.~\ref{sec:constant_motion}. Social amplification therefore affects \emph{which} equilibrium is selected, while the continuum of equilibria itself persists even in the symmetric case
$r_M=r_A$.

Appendix~\ref{appSection:sensitivity} extends this analysis over a broader range of biological and social-amplification parameters.

\section{Discussion}
\label{sec:discussion}

In this work, we have developed a model that illustrates how differences in clinical effectiveness need not translate into a unique population-level treatment share when health outcomes and treatment adoption are dynamically coupled. Although evidence-based medicine is assigned a higher recovery rate $\gamma_M>\gamma_A$, the deterministic dynamics do not generically drive the system toward complete medicine adoption. Instead, the system admits a continuous one-dimensional manifold of equilibria, along which larger medicine shares are associated with lower stationary illness burdens. Clinical superiority therefore determines the ordering of health outcomes along the manifold, but does not by itself determine which population share is ultimately selected.

The exact conserved quantity provides the complementary selection mechanism. For an interior initial condition, the dynamics remain confined to a single invariant level set, which intersects the equilibrium manifold at a unique compatible equilibrium. The resulting neutrality should therefore not be interpreted as unrestricted motion along a continuum of stationary states: different initial health and treatment states may select different equilibria, but each autonomous trajectory is dynamically constrained. Importantly, the existence of the manifold does not require asymmetric social amplification. It persists for $r_M=r_A$ and follows instead from the behavioral coupling: once the treatment-specific illness burdens become stationary, the endogenous social driving term vanishes. The coefficients $r_M$ and $r_A$ influence the transient evolution and hence equilibrium selection, but do not create the continuum itself.

This mechanism differs from standard behavior--disease models, in which behavior commonly feeds back on pathogen transmission and may generate collective effects such as herd protection or free riding \cite{funk2010modelling,fenichel2011adaptive,bauch2004vaccination, wang2016statistical}. Here the focal condition is non-infectious, and the coupling operates instead through treatment choice: changes in the health outcomes associated with each treatment modify its social attractiveness. The framework thus provides a minimal setting in which socially mediated treatment adoption alone generates nontrivial collective dynamics.

Homeopathy serves as a motivating example of the separation between clinical effectiveness and social persistence, but the model is not intended as a quantitative description of homeopathic use in any particular population. The social-amplification coefficients are phenomenological parameters, and the regime $r_A>r_M$ should be understood as a stylized asymmetry rather than as a universal empirical claim. Likewise, the definition of attractiveness through changes in illness burden is a modeling assumption representing an event-sensitive social response. The alternative formulation discussed in Appendix~\ref{appSection:alternativeModel} produces a qualitatively different phase-space structure, illustrating the importance of this behavioral closure without constituting empirical validation of it.

The invariant also clarifies the effect of external perturbations. Direct interventions on treatment adoption can change the invariant level set and therefore leave a persistent displacement after the intervention has ceased. By contrast, perturbations that preserve the conserved quantity may generate large transient excursions without altering the final compatible equilibrium. This suggests a qualitative public-health implication: in the idealized neutral system, temporary interventions can in principle have persistent effects without permanent forcing. The present model, however, does not establish the optimal duration, intensity, or cost effectiveness of such interventions, nor should this persistence be interpreted as evidence that arbitrarily brief campaigns would suffice in real populations.

Several simplifying assumptions delimit these conclusions and provide natural directions for extension. The deterministic mean-field formulation neglects finite-population fluctuations, which may become particularly relevant along a neutrally stable direction and require an explicitly stochastic model to determine whether diffusion, fixation, or other long-time behavior emerges. Network structure, heterogeneous influence, spatial organization, and algorithmically mediated information exposure are likewise absent and could alter both transient dynamics and equilibrium selection. A related simplification concerns treatment switching: the reduced formulation does not resolve whether switching individuals are currently ill or unaffected. Allowing health-dependent switching would require a more disaggregated compartmental model and could introduce additional feedback between health status and treatment choice. Demographic turnover, spontaneous preference changes, forgetting, and institutional influence would also generally break the exact conservation law and may replace perfect neutrality by slow drift toward a smaller set of long-run states. These questions motivate stochastic, networked, and agent-based extensions of the present framework.

In brief, our model provides a minimal analytical baseline in which clinical effectiveness, social feedback, and initial conditions play distinct roles. Superior recovery performance determines which equilibria are biologically preferable, while the coupled dynamics determine which equilibrium is actually selected. The central result is therefore a separation between clinical performance and dynamical selection: even when one treatment is more effective, the long-run population composition may retain a persistent memory of its previous health and treatment state.

\section*{Author contributions}
H.B.-A., A.d.M.-A. and C.G.-L. jointly conceived and developed the study and contributed to the formulation of the mathematical model and the theoretical framework. All authors contributed to the analytical derivations, numerical analysis, interpretation of the results, and critical discussion of the model and its implications. H.B.-A., A.d.M.-A. and C.G.-L. contributed to drafting and revising the manuscript, and all authors approved the final version.



\section*{Use of generative artificial intelligence}
Claude (Anthropic) was used solely to assist with stylistic editing and correction of the English language.

\section*{Data, code and materials accessibility}
The code necessary to reproduce the analyses, together with the processed outputs and figures, is available at \cite{benitoandres2026culturalcompetition}. 

\bibliographystyle{unsrt} 
\bibliography{refs} 

@article{tversky1973availability,
  author  = {Tversky, Amos and Kahneman, Daniel},
  title   = {Availability: A heuristic for judging frequency and probability},
  journal = {Cognitive Psychology},
  volume  = {5},
  number  = {2},
  pages   = {207--232},
  year    = {1973},
  doi     = {10.1016/0010-0285(73)90033-9}
}

@article{astin1998why,
  author  = {Astin, John A.},
  title   = {Why patients use alternative medicine: results of a national study},
  journal = {JAMA},
  volume  = {279},
  number  = {19},
  pages   = {1548--1553},
  year    = {1998},
  doi     = {10.1001/jama.279.19.1548}
}

@article{bauch2004vaccination,
  author  = {Bauch, Chris T. and Earn, David J. D.},
  title   = {Vaccination and the theory of games},
  journal = {Proceedings of the National Academy of Sciences},
  volume  = {101},
  number  = {36},
  pages   = {13391--13394},
  year    = {2004},
  doi     = {10.1073/pnas.0403823101}
}

@article{emprechtinger2022assessing,
  title={Assessing the magnitude of reporting bias in trials of homeopathy: a cross-sectional study and meta-analysis},
  author={Emprechtinger, Michael and Antes, Gerd and Fronczek, Justyna and Stegemann, Miriam and Albrecht, Harald and Grouven, Ulrich and others},
  journal={BMJ Evidence-Based Medicine},
  volume={28},
  number={2},
  pages={101--107},
  year={2022},
  publisher={Institute of Biomedical Ethics and History of Medicine, University of Zurich}
}

@article{shang2005are,
  title={Are the clinical effects of homeopathy placebo effects? Comparative study of placebo-controlled trials of homoeopathy and allopathy},
  author={Shang, Aijing and Huwiler-Müntener, Karin and Nartey, Linda and Jüni, Peter and Dörig, Stephan and Sterne, Jonathan AC and Puhan, Martin A and Egger, Matthias},
  journal={The Lancet},
  volume={366},
  number={9487},
  pages={726--732},
  year={2005}
}

@article{relton2017prevalence,
  author  = {Relton, Clare and Cooper, Kate and Viksveen, Petter and Fibert, Philippa and Thomas, Kate},
  title   = {Prevalence of homeopathy use by the general population worldwide: a systematic review},
  journal = {Homeopathy},
  volume  = {106},
  number  = {2},
  pages   = {69--78},
  year    = {2017}
}

@article{ludtke2008conclusions,
  author  = {L{\"u}dtke, Rainer and Rutten, Alexander L. B.},
  title   = {The conclusions on the effectiveness of homeopathy highly depend on the set of analyzed trials},
  journal = {Journal of Clinical Epidemiology},
  volume  = {61},
  number  = {12},
  pages   = {1197--1204},
  year    = {2008}
}

@article{ernst2002systematic,
  author  = {Ernst, Edzard},
  title   = {A systematic review of systematic reviews of homeopathy},
  journal = {British Journal of Clinical Pharmacology},
  volume  = {54},
  number  = {6},
  pages   = {577--582},
  year    = {2002}
}

@article{centola2010spread,
  title={The spread of behavior in an online social network experiment},
  author={Centola, Damon},
  journal={Science},
  volume={329},
  number={5996},
  pages={1194--1197},
  year={2010},
  publisher={American Association for the Advancement of Science}
}

@article{funk2010modelling,
  title={Modelling the influence of human behaviour on the spread of infectious diseases: a review},
  author={Funk, Sebastian and Salath{\'e}, Marcel and Jansen, Vincent AA},
  journal={Journal of the Royal Society Interface},
  volume={7},
  number={50},
  pages={1247},
  year={2010}
}

@article{berger2011arousal,
  title={Arousal increases social transmission of information},
  author={Berger, Jonah},
  journal={Psychological science},
  volume={22},
  number={7},
  pages={891--893},
  year={2011},
  publisher={Sage Publications Sage CA: Los Angeles, CA}
}

@article{fenichel2011adaptive,
  title={Adaptive human behavior in epidemiological models},
  author={Fenichel, Eli P and Castillo-Chavez, Carlos and Ceddia, M Graziano and Chowell, Gerardo and Parra, Paula A Gonzalez and Hickling, Graham J and Holloway, Garth and Horan, Richard and Morin, Benjamin and Perrings, Charles and others},
  journal={Proceedings of the National Academy of Sciences},
  volume={108},
  number={15},
  pages={6306--6311},
  year={2011},
  publisher={National Academy of Sciences}
}

@article{mathie2014randomised,
  author  = {Mathie, Robert T. and Lloyd, Suzanne M. and Legg, Lynn A. and Clausen, J{\"u}rgen and Moss, Sian and Davidson, Jonathan R. T. and Ford, Ian},
  title   = {Randomised Placebo-Controlled Trials of Individualised Homeopathic Treatment: Systematic Review and Meta-Analysis},
  journal = {Systematic Reviews},
  volume  = {3},
  pages   = {142},
  year    = {2014},
  doi     = {10.1186/2046-4053-3-142}
}

@article{goldman2015social,
  author  = {Goldman, Alyssa W. and Cornwell, Benjamin},
  title   = {Social network bridging potential and the use of complementary and alternative medicine in later life},
  journal = {Social Science \& Medicine},
  volume  = {140},
  pages   = {69--80},
  year    = {2015},
  doi     = {10.1016/j.socscimed.2015.07.003}
}

@techreport{nhmrc2015evidence,
  author      = {{National Health and Medical Research Council}},
  title       = {{NHMRC} Information Paper: Evidence on the Effectiveness of Homeopathy for Treating Health Conditions},
  institution = {National Health and Medical Research Council},
  address     = {Canberra, Australia},
  year        = {2015},
  url         = {https://apo.org.au/node/53563}
}

@article{verelst2016behavioural,
  author  = {Verelst, Frederik and Willem, Lander and Beutels, Philippe},
  title   = {Behavioural change models for infectious disease transmission: a systematic review (2010--2015)},
  journal = {Journal of the Royal Society Interface},
  volume  = {13},
  number  = {125},
  pages   = {20160820},
  year    = {2016},
  doi     = {10.1098/rsif.2016.0820}
}

@article{wang2016statistical,
  author  = {Wang, Zhen and Bauch, Chris T. and Bhattacharyya, Samit and d'Onofrio, Alberto and Manfredi, Piero and Perc, Matja{\v z} and Perra, Nicola and Salath{\'e}, Marcel and Zhao, Dawei},
  title   = {Statistical physics of vaccination},
  journal = {Physics Reports},
  volume  = {664},
  pages   = {1--113},
  year    = {2016},
  doi     = {10.1016/j.physrep.2016.10.006}
}

@article{mathie2017randomised,
  author  = {Mathie, Robert T. and Ramparsad, Nitish and Legg, Lynn A. and Clausen, J{\"u}rgen and Moss, Sian and Davidson, Jonathan R. T. and Messow, Claudia-Martina and McConnachie, Alex},
  title   = {Randomised, Double-Blind, Placebo-Controlled Trials of Non-Individualised Homeopathic Treatment: Systematic Review and Meta-Analysis},
  journal = {Systematic Reviews},
  volume  = {6},
  pages   = {63},
  year    = {2017},
  doi     = {10.1186/s13643-017-0445-3}
}

@article{johnson2018complementary,
  author  = {Johnson, Skyler B. and Park, Henry S. and Gross, Cary P. and Yu, James B.},
  title   = {Complementary Medicine, Refusal of Conventional Cancer Therapy, and Survival Among Patients With Curable Cancers},
  journal = {JAMA Oncology},
  volume  = {4},
  number  = {10},
  pages   = {1375--1381},
  year    = {2018},
  doi     = {10.1001/jamaoncol.2018.2487}
}

@article{lattenaor2018influence,
  author  = {Latte-Naor, Shelly and Sidlow, Robert and Sun, Lingyun and Li, Qing S. and Mao, Jun J.},
  title   = {Influence of family on expected benefits of complementary and alternative medicine (CAM) in cancer patients},
  journal = {Supportive Care in Cancer},
  volume  = {26},
  number  = {6},
  pages   = {2063--2069},
  year    = {2018},
  doi     = {10.1007/s00520-018-4053-0}
}

@article{perra2021non,
  title={Non-pharmaceutical interventions during the COVID-19 pandemic: A review},
  author={Perra, Nicola},
  journal={Physics reports},
  volume={913},
  pages={1--52},
  year={2021},
  publisher={Elsevier}
}

@misc{benitoandres2026culturalcompetition,
  author       = {Benito-Andr{\'e}s, Hermes},
  title        = {Cultural Competition in Medicine},
  year         = {2026},
  howpublished = {\url{https://github.com/hermesBean14/cultural-competition-medicine}},
  note         = {GitHub repository}
}


\clearpage
\newpage
\appendix

\renewcommand{\thefigure}{S\arabic{figure}}
\setcounter{figure}{0}

\begin{center}
    \LARGE\bfseries Supplementary Information
\end{center}

\vspace{1cm}

\section{Methodological considerations: health trends vs. absolute recoveries}
\label{appSection:alternativeModel}

In formulating the behavioral replicator dynamics, the definition of social attractiveness is intended to represent an event-sensitive response to observed health outcomes. According to this formulation, individuals may become more favorable toward a treatment when they perceive a net positive momentum in health outcomes. The social signal is not driven merely by the instantaneous recovery flow; it is also reduced by new incidence. When adopters of a treatment fall ill, the net health trend becomes less favorable, reducing the treatment's social attractiveness. Therefore, the social attractiveness of a treatment is assumed to depend on the net reduction of its illness burden (recoveries minus new cases), making $\alpha_k = - r_k \dot{I}_k$ the baseline formulation considered in the main text.

To illustrate the dynamical consequences, one might consider an alternative formulation that equates attractiveness directly to the instantaneous recovery flow, $\alpha_k = r_k \gamma_k I_k$, completely ignoring the dampening effect of new illnesses. Under this assumption, the replicator equation becomes:
\begin{equation}
    \frac{dM}{dt} = M(1 - M) \left[ r_M \gamma_M I_M - r_A \gamma_A I_A \right]\, .
     \label{eq:alternativeModel}
\end{equation}

Evaluating this alternative system at the biological steady state---where the endemic burdens scale as $I_M^* = M^* \beta / (\beta + \gamma_M)$ and $I_A^* = (1 - M^*) \beta / (\beta + \gamma_A)$---reveals a qualitative change in the equilibrium structure. For strictly positive parameters, the one-dimensional equilibrium manifold is destroyed, and the system instead exhibits a bistable regime with three isolated fixed points: two stable boundaries ($M^* = 0$ and $M^* = 1$) and an unstable interior saddle.

\begin{figure*}[htbp]
\centering
\includegraphics[width=\linewidth]{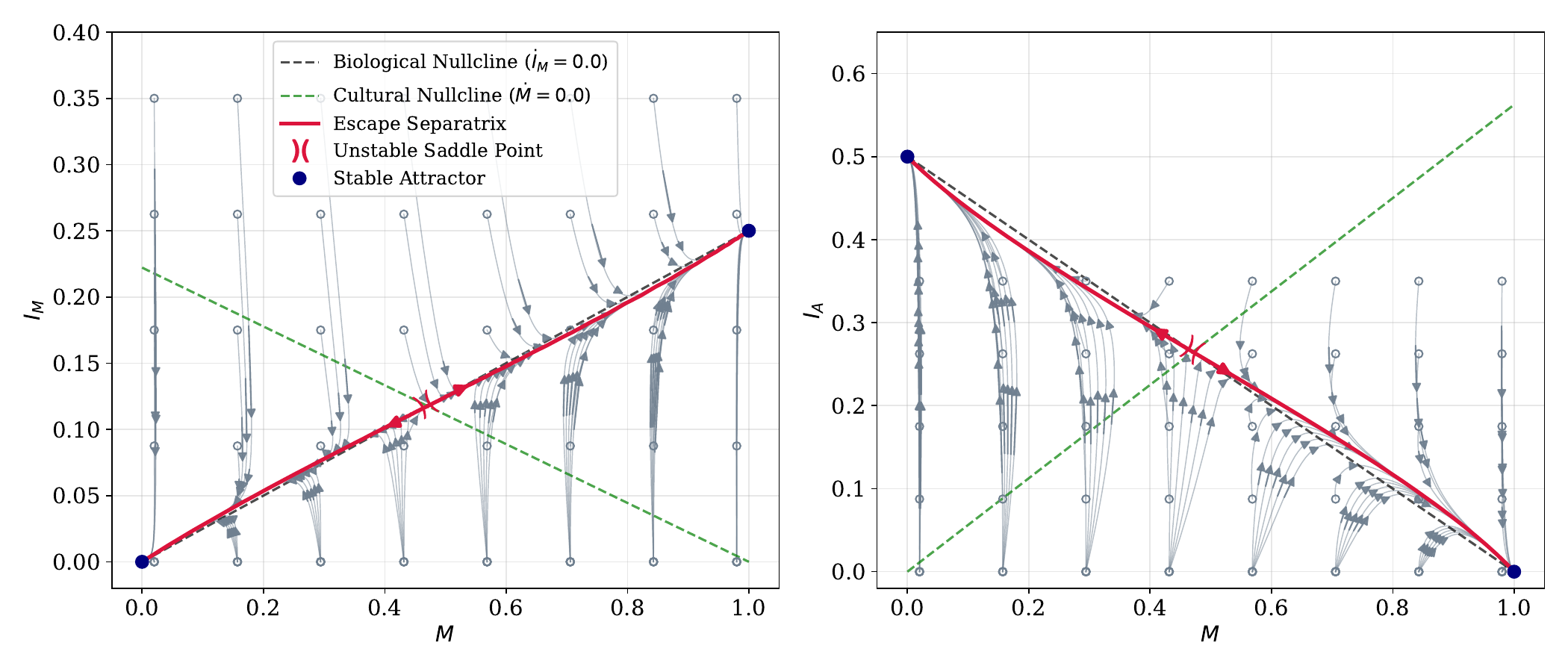}
    \caption{
        \textbf{Topological Collapse (Alternative Model)}. Two projections of the phase space, $M$ vs.\ $I_M$ (panel a) and $M$ vs.\ $I_A$ (panel b). These phase portraits illustrate the basin of attraction for the alternative absolute-recovery model depicted in Eq.~\eqref{eq:alternativeModel}. Open circles denote the initial conditions for the simulated trajectories (thin lines), which flow towards the stable attractors at the extreme polarization corners ($M = 0$ and $M = 1$), indicated by solid circles. The unstable interior saddle point (marked with a saddle symbol) repels the flow, with the exact escape separatrices highlighted as thick lines. This dynamics contrasts with the coexistence manifold shown in Figure 2 of the main text. Parameters used: $\beta = 0.2$, $\gamma_M = 0.6$, $\gamma_A = 0.2$, $r_M = 1$, $r_A = 1.333$.}
\label{fig:alternativeModelPhaseDiagram}
\end{figure*}

Figure \ref{fig:alternativeModelPhaseDiagram} maps this fragmented architecture across two phase space projections ($M$ vs.\ $I_M$ in panel a, and $M$ vs.\ $I_A$ in panel b). The visualized flow dynamics clearly illustrate the role of the unstable interior saddle as a watershed: it repels all trajectories towards the stable fixed points at the extreme boundaries, effectively splitting the domain into two mutually exclusive basins of attraction.

At the population level, this alternative bistable topology produces a \textit{winner-takes-all} outcome within the model, in which one treatment eventually dominates. This qualitative behavior contrasts sharply with the continuous coexistence supported by the derivative-based formulation. The comparison therefore highlights the structural importance of the chosen behavioral closure: replacing the net change in illness burden by the instantaneous recovery flow removes the continuous equilibrium manifold and replaces path-dependent equilibrium selection by bistability between the two boundary states. The continuous coexistence generated by the derivative-based formulation is also qualitatively more consistent with the empirical persistence of alternative medicine alongside evidence-based medical practice than the strict winner-takes-all outcome predicted by the absolute-recovery formulation.

\begin{figure*}[htbp] 
\centering 
\includegraphics[width=0.6\linewidth]{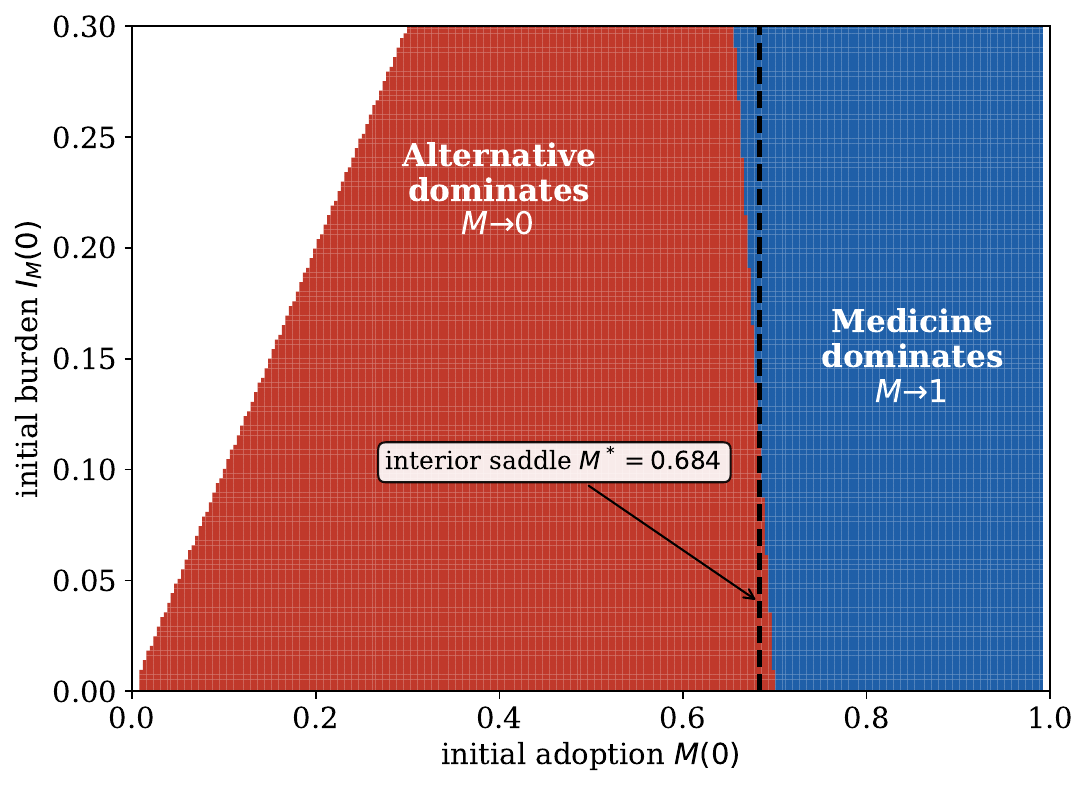} 
\caption{
    \textbf{Basin of attraction for the alternative absolute-recovery model.} The heatmap displays the asymptotic outcome of the system initialized across a grid of starting adoption levels $M(0)$ and disease burdens $I_M(0)$, with $I_A(0)$ initialized at its respective biological equilibrium. The alternative formulation ($\alpha_k = r_k \gamma_k I_k$) destroys the continuous manifold, yielding strict bistability. The dashed line indicates the unstable interior saddle ($M^* \approx 0.684$), which acts as a watershed separating the basins: trajectories collapse either to complete alternative-medicine dominance (red region, $M \to 0$) or full evidence-based medicine adoption (blue region, $M \to 1$). Fixed parameters: $\beta = 0.1$, $\gamma_M = 0.5$, $\gamma_A = 0.15$, $r_M = 1$, $r_A = 3$.} 
\label{fig:SI_1} 
\end{figure*}

To further display this topological collapse, we visualize the basins of attraction generated by this alternative formulation. Figure \ref{fig:SI_1} maps a grid of initial conditions—spanning initial evidence-based medicine adoption $M(0)$ and initial evidence-based medicine-associated illness burden $I_M(0)$—colored by their final equilibrium state. The presence of the interior saddle point ($M^* \approx 0.684$) acts as a boundary between the two basins of attraction, replacing the continuous manifold by two distinct asymptotic domains. Trajectories starting to the left of the saddle inevitably fall into the alternative-medicine-dominant attractor (red region, $M \to 0$), while those starting to the right are pulled toward full evidence-based medicine adoption (blue region, $M \to 1$). This \textit{winner-takes-all} bistability illustrates how replacing the derivative-based coupling by an absolute-recovery coupling removes the model's continuous coexistence structure.

\section{Extended sensitivity analysis and parameter space}
\label{appSection:sensitivity}

To complement the sensitivity analysis presented in the main text, this section provides a broader visualization of the parameter space and the equilibrium structure of the continuous manifold of equilibria.

\begin{figure*}[htbp] 
\centering \includegraphics[width=\linewidth]{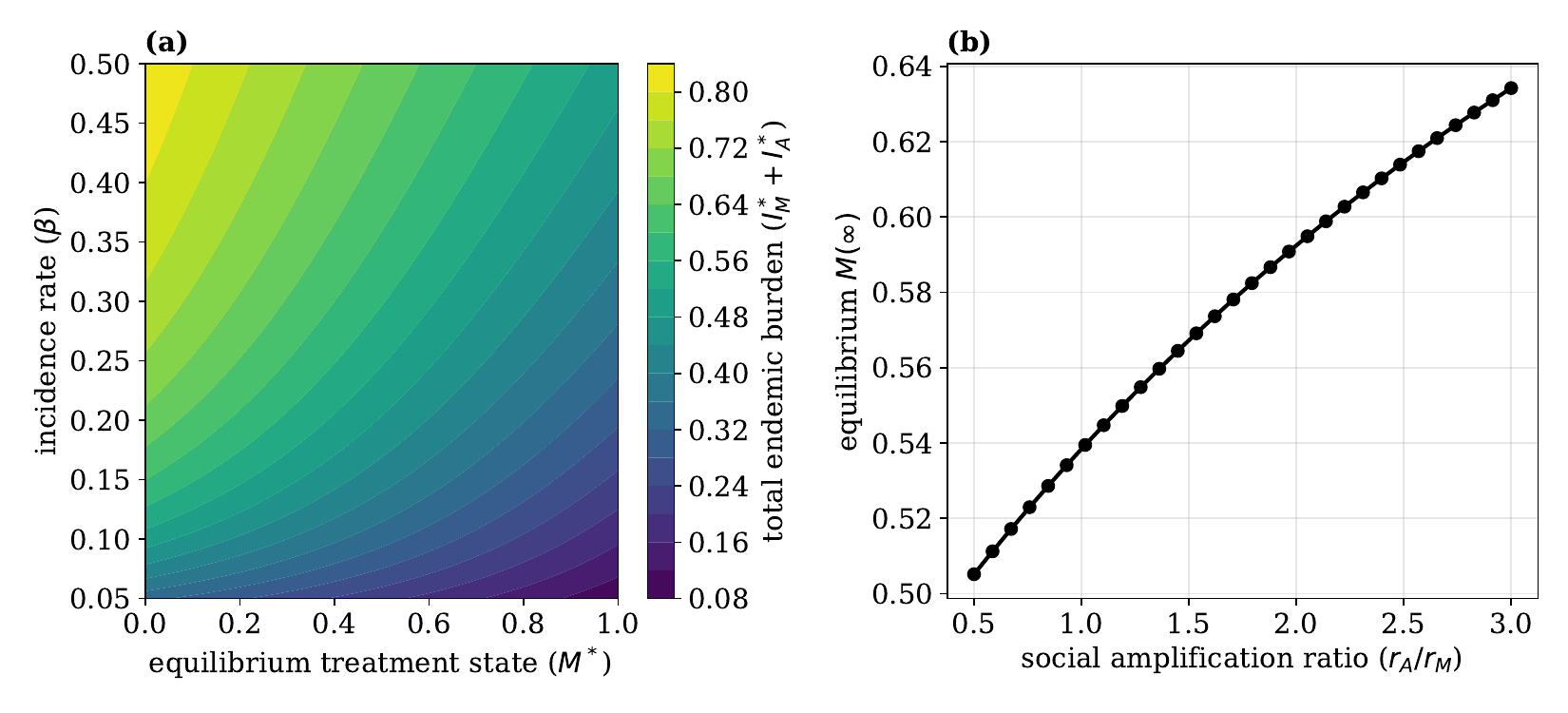} 
\caption{
    \textbf{Extended Sensitivity Analysis.} \textbf{(a)} Analytical contour map (heatmap) showing the total equilibrium illness burden ($I_M^* + I_A^*$) as a function of the incidence rate $\beta$ and the equilibrium treatment share $M^*$ along the continuous manifold. \textbf{(b)} Numerical sweep of the selected equilibrium treatment share $M_\infty$ against the social-amplification ratio $r_A/r_M$. The system is integrated using a Runge-Kutta method (RK45) from an initially healthy out-of-equilibrium state, $\vec{X}(0) = (0.5, 0, 0)$, illustrating the dependence of equilibrium selection on social amplification predicted by the analytical selection relation. Fixed parameters for both the analytical bounds and the numerical integration are $\gamma_M = 0.5$ and $\gamma_A = 0.1$. Additionally, panel b uses $\beta = 0.2$ and $r_M = 1$.} 
\label{fig:SI_sensitivity} 
\end{figure*}

Figure \ref{fig:SI_sensitivity}a presents an analytical contour map of the total equilibrium illness burden, $I_{\mathrm{total}}^* = I_M^* + I_A^*$, evaluated across the entire continuous manifold ($M^* \in [0, 1]$) and a wide range of incidence rates ($\beta \in [0.05, 0.5]$). This heatmap visually reinforces that for any given incidence rate, shifting the population towards evidence-based medicine ($M^* \to 1$) consistently minimizes the macroscopic burden, given the underlying clinical asymmetry ($\gamma_M > \gamma_A$). The smooth gradient is consistent with the continuous variation of the equilibrium illness burden along the manifold.

Figure \ref{fig:SI_sensitivity}b examines the relaxation from an initially healthy population. Here, the numerical integration is strictly initialized from a completely healthy population ($\vec{X}(0) = (0.5, 0, 0)$). By sweeping the social-amplification ratio $q=r_A / r_M$, we observe a monotonic increase in the selected evidence-based medicine share $M_\infty$. As discussed in the main text, because the behavioral replicator evaluates the net flux of illness, an initially healthy population guarantees a transient governed by new illness ($\dot{I}_k > 0$). Consequently, increasing $q$ amplifies the negative contribution associated with the initially increasing alternative-medicine-related illness burden and shifts the compatible equilibrium toward a larger evidence-based medicine share. This behavior is precisely the sign predicted by Eq.~\eqref{eq:sensitivity_rA} for $I_{A,0}=0$.

\end{document}